\documentclass[aps,prl,twocolumn,showpacs,superscriptaddress]{revtex4}
\usepackage{amssymb}
\usepackage{color}
\usepackage{amsmath}
\usepackage{graphicx}
\usepackage{bm}
\usepackage{float} 
 \usepackage{natbib} 
\begin{document}

\title{D-carbon: A New \textit{sp}$^3$ Carbon Allotrope}

\author{Dong Fan} 
\affiliation{College of Materials Science and Engineering, Zhejiang University of Technology,Hangzhou 310014, China} 

\author{Shaohua Lu}
\affiliation{College of Materials Science and Engineering, Zhejiang University of Technology,Hangzhou 310014, China} 

\author{Xiaojun Hu} 
\affiliation{College of Materials Science and Engineering, Zhejiang University of Technology,Hangzhou 310014, China} 

\begin{abstract}
We have investigated the structural, mechanical and electronic properties of D-carbon by  $\textit{ab initio}$ calculations, a new phase of crystalline \textit{sp}$^3$ carbon (pace group D$_{2h}$$^5$, $\textit{Pmma}$-orthorhombic). Total-energy calculations demonstrate that D-carbon is energetically more favorable than previously proposed T$_6$ structure. State-of-the-art theoretical calculations show that this new phase is dynamic, mechanical, and thermal stable at zero pressure, and more stable than graphite beyond 63.7 GPa. Importantly, the calculations reveal that D-carbon possesses high Vickers hardness (86.58 GPa) and bulk modulus (369 GPa), which are comparable to diamond. D-carbon is a semiconductor with a band gap of 4.33 eV, lower than diamond's gap (5.47 eV).  The simulated X-ray diffraction pattern is in satisfactory agreement with the previously experimental data in chimney or detonation soot, suggesting its possible presence in the specimen. Equally important, the possible transition path from diamond to D-carbon has been investigated, indicating a possible approach to synthesize this new phase.
\end{abstract}


\maketitle

 Carbon is an amazing and versatile element: not only because it is the significant and essential element required for all life processes, but also, due to its rich physical and chemical properties. The discovery of fullerenes,\cite{kroto1985c60} nanotubes,\cite{iijima1993single} and graphene\cite{novoselov2004electric} has motivated tremendous interest in recent years to explore newly carbon structures in \textit{sp}$^3$-, \textit{sp}$^2$-, and \textit{sp}-hybridized bonding networks. These synthesized carbon allotropes give rise to enormous scientific and technological impacts on natural science, leading to many applications in different fields such as protective coatings, gas sensing, energy storage systems, and solar cells.\cite{balandin2011thermal,lee2006recent} On the other hand, to guide experimental design, highly accurate theoretical predictions are indispensable in the search of new carbon allotropes. Various elusive carbon allotropic modifications have been proposed, including the Cco C$_8$,\cite{zhao2011novel} bco C$_{16}$,\cite{wang2016body} bcc C$_8$,\cite{johnston1989superdense} T-carbon,\cite{sheng2011t} T$_6$-, and T$_{14}$-carbon.\cite{zhang2013stable} Several predicted carbon phases, e.g., monoclinic bct C$_4$,\cite{umemoto2010body} and M-carbon,\cite{li2009superhard} were also proposed to simulate the synthesized phase. And it's worth noting that the previously theoretical predicted T-carbon has been synthesized very recently,\cite{zhang2017pseudo} although this structure is  thermodynamically metastable phase with the high total energy (-7.92eV/atom).\cite{sheng2011t} Moreover, there are still other experimental and theoretical efforts made on new carbon materials (e.g., V carbon\cite{yang2017novel} and compressed glassy carbon\cite{hu2017compressed}). Therefore, it is more significant to enter into the era of carbon allotropes.\cite{hirsch2010era}

In this letter, on the basis of the first-principles calculations, we present a theoretical investigation of structural, mechanical and electronic properties of the new carbon allotrope, D-carbon, namely. The calculated results demonstrate that D-carbon, dynamically stable and with a lower energy than previously reported modifications (e.g., C$_{20}$, T$_6$, and T-carbon), has a Vickers hardness 86.58 GPa mildly smaller than diamond (93.7 GPa). A satisfactory match of simulated and measured X-ray diffraction pattern indicates the possible presence of D-carbon in chimney or detonation soot. 

\begin{figure}[htbp]
\centering
\includegraphics[width=0.8\columnwidth]{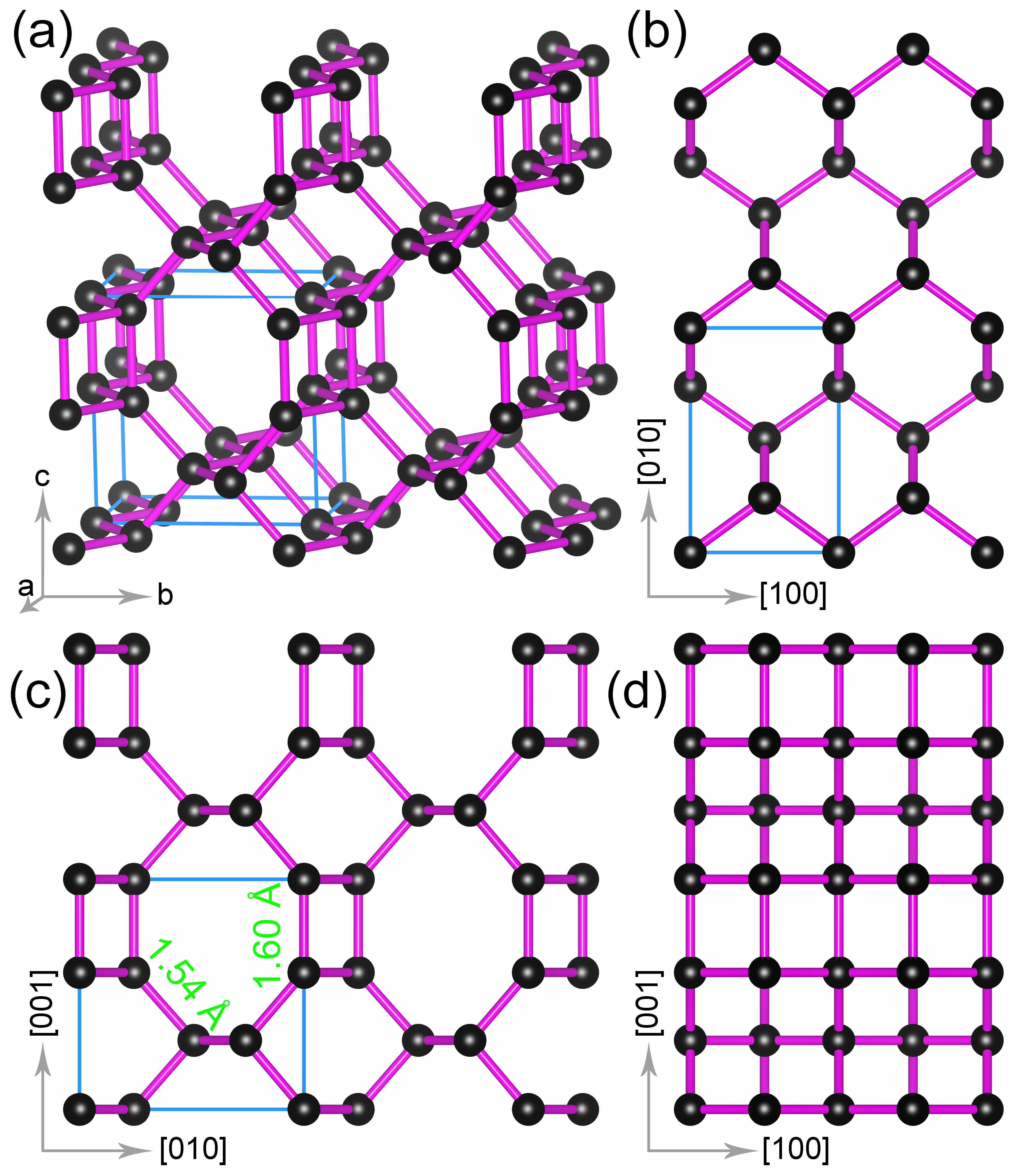}
\caption{Crystal structure of D-carbon: (a) perspective view, (b-d) view along the $\bm{c}$,  $\bm{a}$, and  $\bm{b}$ axis, respectively. It has a 6-atoms tetragonal structure with lattice parameters $\bm{a}$ = 2.52 \r{A}, $\bm{b}$ = 3.81 \r{A}, and $\bm{c}$ = 3.91 \r{A}. The carbon atoms are plotted with black balls.}
\end{figure}

The first-principles calculations were based on density functional theory with generalized gradient approximation (GGA) in the form of Perdew-Burke-Ernzerhof function for exchange-correlation potential.\cite{perdew1996generalized} All the calculations were performed using the Vienna $\textit{ab initio}$ Simulation Package (VASP).\cite{kresse1996efficient} The energy cutoff of the plane wave was set to 650 eV with the energy precision of 10$^{-5}$ eV.\cite{fan2017novel} The atomic positions were fully relaxed until the maximum force on each atom was less than 10$^{-3}$ eV/\r{A}. For a carbon atom, 2$\textit{s}$$^2$2$\textit{p}$$^2$ electrons were considered as the valence electrons. The Brillouin zone was sampled with a 10 $\times$ 7 $\times$ 6 Monkhorst-Pack k-points grid for geometry optimization. Phonon dispersions and frequency densities of states (DOS) were performed in the Phonopy package\cite{togo2015first} interfaced with the density functional perturbation theory (DFPT)\cite{baroni2001phonons} as performed in VASP. For accurate bandgap estimations, we employed the hybrid functional approach (HSE06).\cite{heyd2003hybrid} First-principles finite temperature molecular dynamics (MD) simulations were performed to further examined the stability of the structure by using time steps of 1 femtosecond in 3 $\times$ 3 $\times$ 3 super-cells containing 162 atoms/cell.

Fig. 1 shows the optimized structural model of D-carbon. This structure has a orthorombic primitive cell containing six C atoms, with a high symmetric space group $\textit{Pmma}$ (D$_{2h}$$^5$, 51). At zero pressure, the relaxed bond lengths of C-C are, respectively, 1.60 and 1.54 \r{A}. In contrast to the uniform bond length of 1.54 \r{A} in diamond, this bonding is quite unusual, notably longer than the bond length of 1.42 \r{A} in graphene, but slightly shorter than previously reported longest C-C bond (1.788 \r{A}\cite{lu2014self}). 

\begin{figure}[htbp]
\centering
\includegraphics[width=0.9\columnwidth]{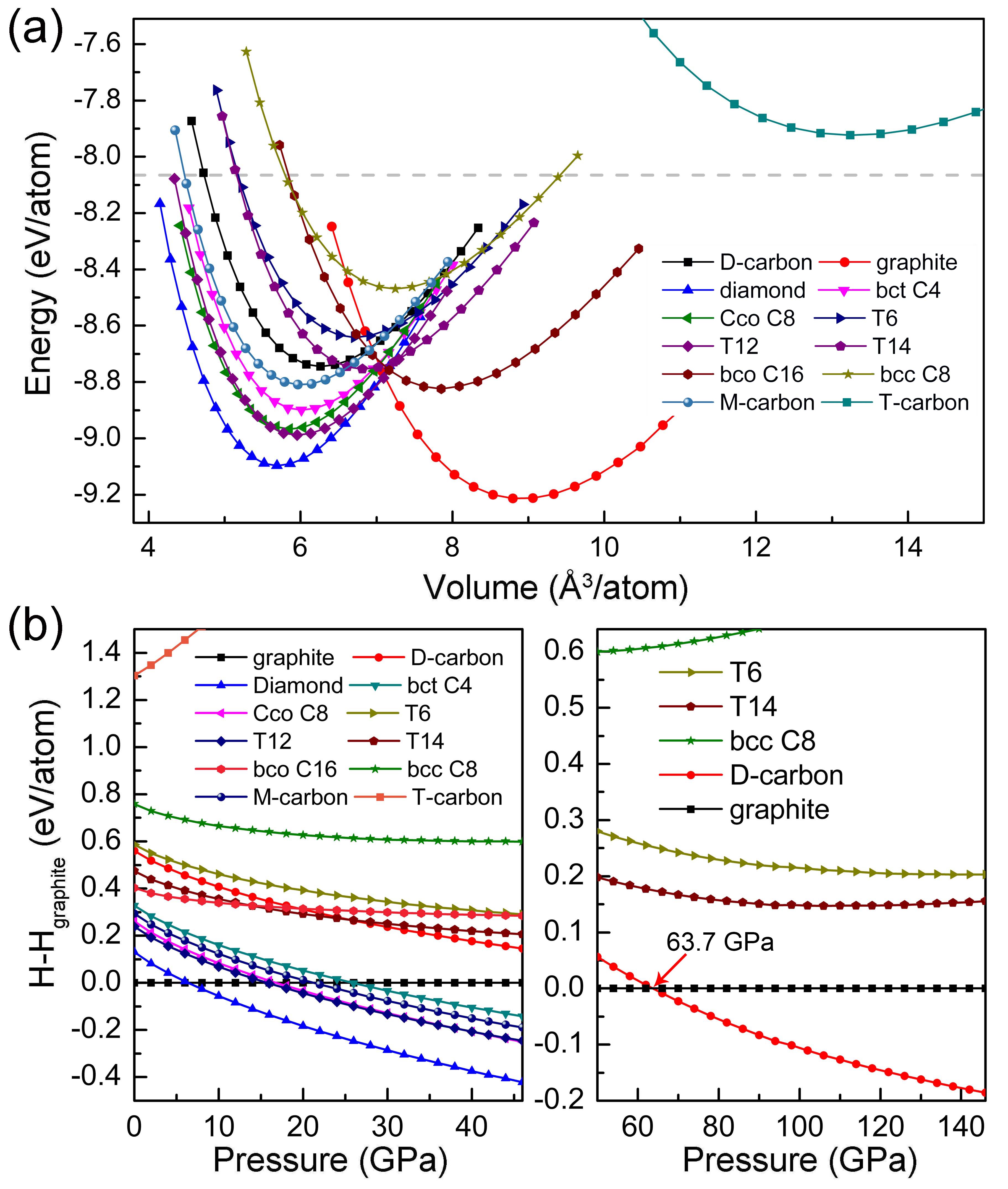}
\caption{(a) Calculated energy versus volume per atom for D-carbon structure compared to graphite, diamond, bct C$_4$, Cco C$_8$, T$_6$, T$_{14}$, bco C$_{16}$, bcc C$_8$, M-carbon, and T-carbon. The dashed line indicates the energy level of C$_{20}$ fullerene.\cite{prinzbach2000gas} (b) Calculated relative enthalpies of D-carbon, diamond, bct C$_4$, Cco C$_8$, T$_6$, T$_{14}$, bco C$_{16}$, bcc C$_8$, M-carbon, and T-carbon with respect to graphite.}
\end{figure}

Fig. 2 shows the calculated total energy versus volume and relative enthalpy for D-carbon compared to other previously proposed carbon phases. We note that D-carbon is not only more stable than some theoretically predicted carbon modifications (e.g., T$_6$, bcc C$_8$), but also energetically more favorable than several experimentally realized carbon modifications (e.g., C$_{20}$ fullerene and T-carbon), implying that the D-carbon could be synthesized. To further evaluate the relative stability of this new phase, we also calculated its cohesive energy $E_{coh} = [6E_{C}-E_{total}]/6$, where $E_{total}$ and $E_{C}$ are the total energies of D-carbon and a single C atom, respectively. The calculated cohesive energy (7.48 eV/atom), apparently higher than recently experimental synthesized T-carbon,(6.573eV/atom\cite{sheng2011t,zhang2017pseudo}) suggests that D-carbon is a strongly bonding network. Significantly, with increase of the pressure, the D-carbon becomes preferable to graphite above 63.7 GPa and is more stable than earlier theoretical T$_6$ and T$_{14}$ structures at this pressure. (see Fig. 2b) Further, phonon spectrum and (Ph DOS) density of states indicate that D-carbon is dynamically stable, both at zero pressure and high pressure. (see Fig. S1) Therefore, once synthesized, D-carbon should be quenchable as a metastable phase to ambient pressure and low temperatures. The thermal stability of D-carbon was also confirmed by analyzing the backbone root-mean-square deviation (RMSD) from the starting crystal configuration over the process of the trajectory, it is evident that the RMSD levels off to $\sim$3.7 \r{A} at 1000 K, indicating that the geometric configuration is expected to be remarkably stable.

\begin{figure}[htbp]
\centering
\includegraphics[width=0.99\columnwidth]{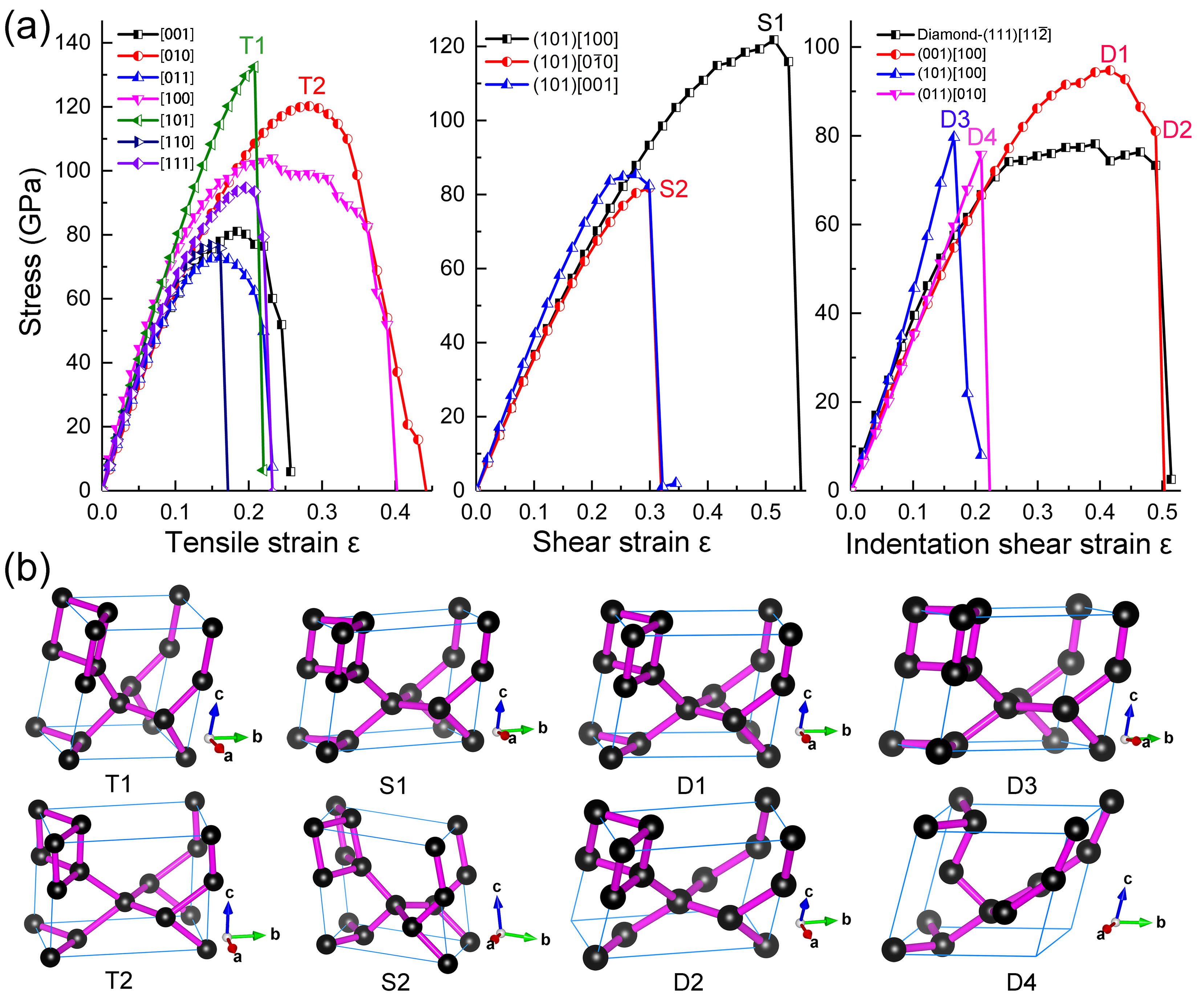}
\caption{(a) Calculated stress responses under pure tensile,(left) pure shear,(center) and indentation shear(right) strains along various high-symmetry directions, respectively. (b) The structural snapshots at different key points after the large deformation of stress on each stress-strain curve under pure tensile, pure shear, or indentation shear strains.}
\end{figure}

D-carbon has nine independent elastic constants C$_{ij}$ and for a stable orthorhombic structure its corresponding elastic constants C$_{ij}$ should satisfy the following elastic stability criteria: $C_{11}, C_{22}, C_{33}, C_{44}, C_{55}$, and $C_{66} > 0$, $[C_{11}+C_{22}+C_{33}+2(C_{12}+C_{13}+C_{23})]>0$, $[C_{11}+C_{22}-2C_{12}]>0, [C_{11}+C_{33}-2C_{13}]>0$, and $[C_{22}+C_{33}-2C_{23}]>0$ for orthorhombic phase.\cite{wang2016body} The calculated elastic constants, C$_{11}$, C$_{22}$, C$_{33}$, C$_{44}$, C$_{55}$, C$_{66}$, C$_{12}$, C$_{13}$, and C$_{23}$ are 1036, 737, 867, 290, 371, 415, 48, 59, and 236 GPa, respectively. Distinctly, the calculated elastic constants meet this criteria, indicating that it is mechanically stable.  As the density of a material is critical for its mechanical properties. The density of D-carbon is 3.18 g/cm$^3$ , which is comparable to that of bct C$_4$ (3.31 g/cm$^3$), but still drastically higher than T-carbon (1.50 g/cm$^3$). Vickers hardness ($H_{\nu}$) of D-carbon is 86.58 GPa, which is calculated by the empirical formula $H(GPa)=350[(N_e^{2/3})e^{-1.191f_i}]/d^{2.5}$,\cite{gao2003hardness} where $N_e$ is the electron density of the number of valence electrons per cubic angstroms, $d$ is the bond length, and $f_i$ is the ionicity of the chemical bond in a crystal scaled by Phillips.\cite{phillips1970ionicity} The calculated Vickers hardness of diamond in this work is 92.49 GPa,\cite{SI} closing to the  previously  theoretical (93.6 GPa) and  experimental value (96$\pm$5 GPa). According to the generally accepted convention, a superhard material owns $H_{\nu}$ $>$ 40 GPa,\cite{solozhenko2001synthesis}  therefore, D-carbon can be regarded as superhard material.

\begin{figure}[htbp]
\centering
\includegraphics[width=0.8\columnwidth]{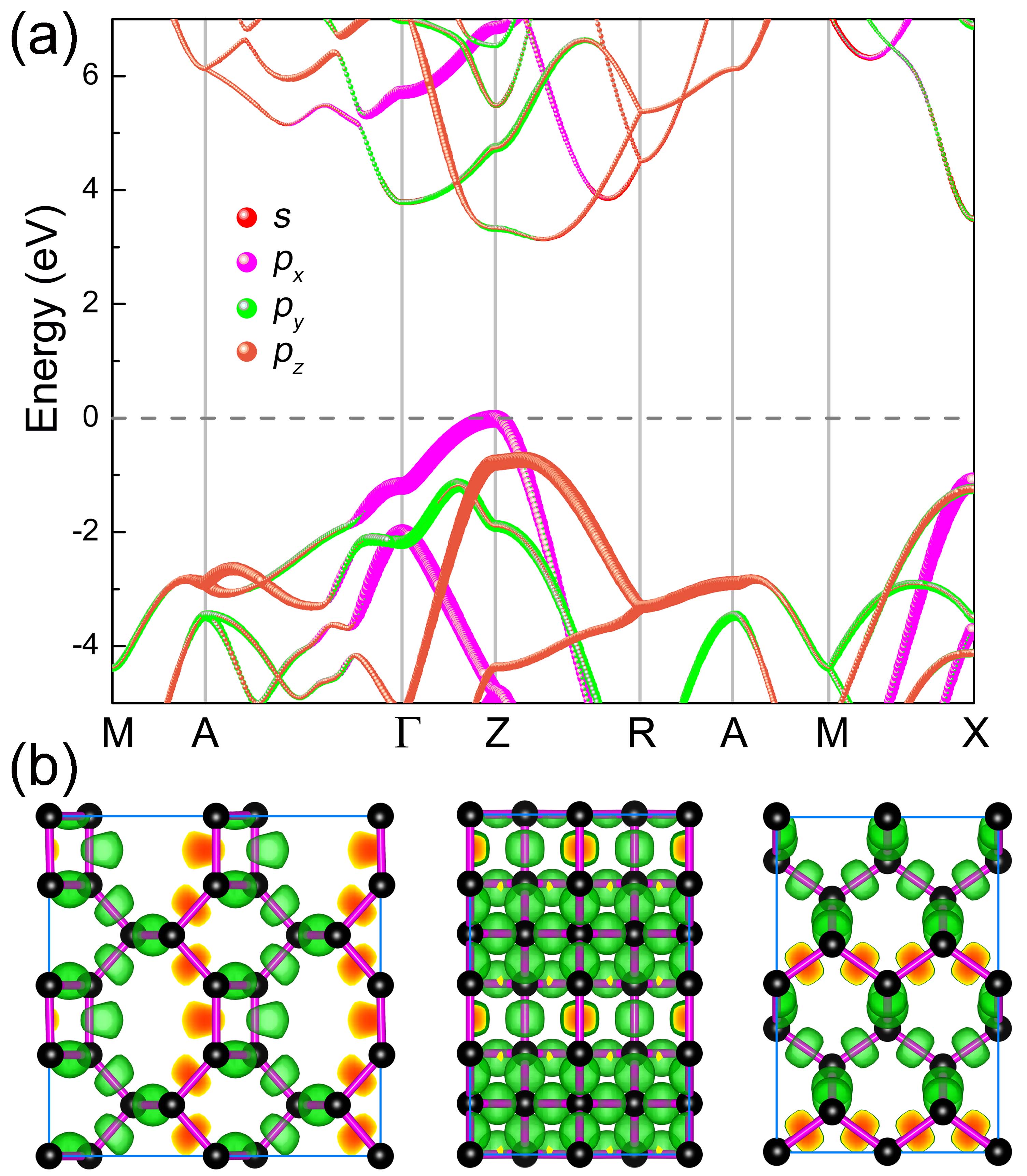}
\caption{(a) Electronic band structure of D-carbon decomposed with respect to the C$_s$, C$_{px}$, C$_{py}$, and C$_{pz}$ orbitals calculated using PBE functional. (b) Calculated ELF of D-carbon at 0 GPa. The isosurface value is set as 0.75.}
\end{figure}

Stress-strain curves for D-carbon under various strains are shown in Fig. 3a. The ideal strength of D-carbon is strongly anisotropic, and interestingly, tensile stress-strain curves along [010], [100], and [011] show a plastic deformation relation, which is consistent with the ductility of D-carbon compared with diamond. D-carbon with [101] orientation presents the largest ideal strength of 133 GPa and critical strain reaches to 0.21. However, with [011] direction, D-carbon shows the smallest ideal strength and critical strain (73 GPa and 0.16), giving rise the (001) easy cleavage planes. In the (101) lattice plane, its pure shear stress along the [100] direction has the highest peak value (122 GP), which is moderately lower than (111)[$\bar{1}$$ \bar{1}$2] direction of diamond (140 GPa),\cite{zhang2006structural} but still obviously higher than $(100)<001>$ slip orientation of T-carbon (7.3 GPa)\cite{chen2011hardness}, while the [0$\bar{1}$0] direction has the lowest peak value (xxGPa). Under indentation shear deformation, the stress response is nearly identical to that under pure shear deformation, but pure shear strain causes greatly enhanced stiffness.

Fig. 4a shows the orbitally resolved band structure of D-carbon. It is insulating with an indirect band gap. The PBE band gap is 3.15 eV. The HSE correction does not change the band structure qualitatively, but increases the band gap to 4.33 eV,(see Fig. S2)\cite{SI} which is smaller than diamond. The results show that the valence band maximum (VBM) of D-carbon is contributed by C$_{2px}$ states while the conduction band minimum (CBM) is mainly contributed by C$_{2pz}$ states. There is obvious orbital hybridization between C$_{2s}$ and C$_{2p}$ states below the Fermi level. To investigate the bonding state between the C-C atoms, we calculated its electron localization function (ELF). From Fig. 4b, the D-carbon is an all-\textit{sp}$^3$ carbon modification with all electrons well localized between the C-C atoms forming the $\sigma$ bonds.

\begin{figure}[htbp]
\centering
\includegraphics[width=0.9\columnwidth]{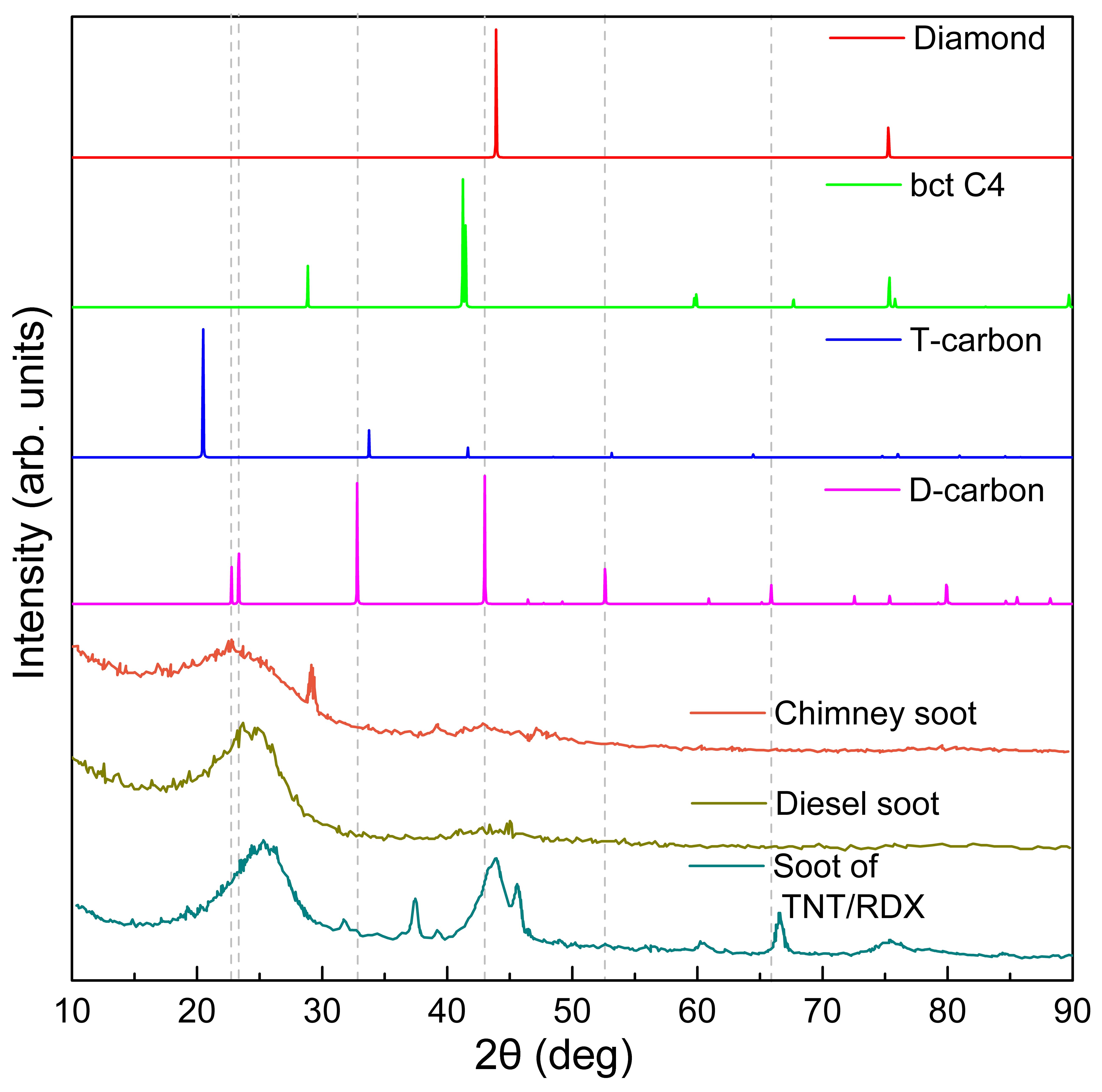}
\caption{The simulated XRD patterns of diamond, bct C$_4$, T-carbon, and D-carbon, compared with experimental data. The experimental XRD patterns from Ref. \cite{chen2003characterization} and Ref. \cite{pantea2006morphological}. The used X-ray wavelength is 1.54 \r{A} as employed in the experiment.\cite{pantea2006morphological} }
\end{figure}

To further establish the experimental connection of our proposed structure, we have simulated the X-ray diffraction (XRD) patterns to compare with the experimental data. (Fig. 5) Different from diamond where the peaks of (111) at 44$^{\circ}$ are observed, for D-carbon, the peaks at 23.3$^{\circ}$, 32.8$^{\circ}$, and 43.1$^{\circ}$ are strong intensity, and one peak at 22.7$^{\circ}$ with weak intensity is also observed. For chimney or detonation soot, the most distinct feature of the experimental measured XRD spectra is the peak around 23$^{\circ}$ that does not match any previously known carbon phases.\cite{chen2003characterization,pantea2006morphological} Our simulated XRD results show that the diffraction peak of D-carbon satisfactory matches the previously unexplained peak, even though the peaks are broad. These results suggest that D-carbon is a possible candidate of the new carbon phase observed in the chimney or detonation soot.

The simulated Raman and IR vibrational modes with corresponding frequencies are presented in Fig. S3. The Raman and IR spectra exhibit distinguishing lines at 1075 cm$^{-1}$ and 1239 cm$^{-1}$, respectively. All these attainable features may be helpful for identifying the D-carbon experimentally. Additionally, transition path simulation reveals the formation of D-carbon through diamond exists a relatively high energy barrier of 0.94 eV/atom, indicating that the D-carbon can be preserved once formed in experiments.(see Fig. S4, \cite{SI})

In summary, the D-carbon, an orthorhombic \textit{sp}$^3$ bonded structure, was discovered theoretically using the first-principles calculations.This structure is energetically more stable than several previously experimental or theoretical \textit{sp}$^3$ species. The results in this Letter indicate that D-carbon is another potential modification of metastable \textit{sp}$^3$ carbon system.

This work was supported by the National Natural Science Foundation of China (Grant Nos. 11504325, 50972129, and 50602039), and Natural Science Foundation of Zhejiang Province (LQ15A040004). This work was also supported by the international science technology cooperation program of China (2014DFR51160), the National Key Research and Development Program of China (No. 2016YFE0133200), the European Union's Horizon 2020 Research and Innovation Staff Exchange (RISE) Scheme (No. 734578), and the One Belt and One Road International Cooperation Project from Key Research and Development Program of Zhejiang Province.

\bibliographystyle{unsrt}
\bibliography{}

\clearpage

\end{document}